\documentclass[aps,pra,twocolumn,superscriptaddress,eps2pdf]{revtex4-1}

\usepackage{epsfig}
\usepackage{amsfonts}
\usepackage{amssymb}
\usepackage{mathrsfs}
\usepackage{theorem}
\usepackage{amsmath}
\usepackage{array,float}
\usepackage{longtable}

\usepackage{subfigure}

\newcolumntype{L}{>{\centering\arraybackslash}m{3cm}}

\usepackage{ifpdf}
\ifpdf
\usepackage{epstopdf}   
\fi

\theoremstyle{plain}

\newcommand{\ket}[1]{\ensuremath{\vert#1\rangle}}

\newcommand{\kb}[2]{\ensuremath{\vert #1 \rangle \langle #2 \vert}}

\renewcommand{\vec}[1]{\ensuremath{\mathbf{#1}}}

\def\id{\mbox{\small 1} \!\! \mbox{1}}

\def\id{{\mathchoice {\rm 1\mskip-4mu l} {\rm 1\mskip-4mu l} {\rm 1\mskip-4.5mu l} {\rm 1\mskip-5mu l}}}
\begin{document}

\title{A random compiler for fast Hamiltonian simulation}

\author{Earl Campbell}

\affiliation{%
 Department of Physics and Astronomy, University of Sheffield, Sheffield, UK
}%

\date{\today}

\begin{abstract}
The dynamics of a quantum system can be simulated using a quantum computer by breaking down the unitary into a quantum circuit of one and two qubit gates. The most established methods are the Trotter-Suzuki decompositions, for which rigorous bounds on the circuit size depend on the number of terms $L$ in the system Hamiltonian and the size of the largest term in the Hamiltonian $\Lambda$.  Consequently, Trotter-Suzuki is only practical for sparse Hamiltonians.  Trotter-Suzuki is a deterministic compiler but it was recently shown that randomised compiling offers lower overheads.  Here we present and analyse a randomised compiler for Hamiltonian simulation where gate probabilities are proportional to the strength of a corresponding term in the Hamiltonian.  This approach requires a circuit size independent of $L$ and $\Lambda$, but instead depending on $\lambda$ the absolute sum of Hamiltonian strengths (the $\ell_1$ norm).  Therefore, it is especially suited to electronic structure Hamiltonians relevant to quantum chemistry.  Considering propane, carbon dioxide and ethane, we observe speed-ups compared to standard Trotter-Suzuki of between $306\times$ and $1591\times$ for physically significant simulation times at precision $10^{-3}$.  Performing phase estimation at chemical accuracy, we report that the savings are similar. 
\end{abstract}

\maketitle

Quantum computers could be used to mimic the dynamics of other quantum systems, providing a computational method to understand physical systems beyond the reach of classical supercomputers. A quantum computation is broken down into a discrete sequence of elementary one and two qubit gates.  To simulate the continuous unitary evolution of the Schr\"{o}dinger equation, an approximation must be made into a finite sequence of discrete gates.  The precision of this approximation can be improved by using more gates.  The standard approaches are the Trotter and higher order Suzuki decompositions~\cite{suzuki1990fractal,suzuki1991general,berry2007efficient}.   In addition to simulating dynamics, we are often interested in learning the energy spectra of Hamiltonians.  Assuming a good ansatz for the ground state, we can combine quantum simulation with phase estimation to find the energy of the ground state~\cite{abrams1999quantum} and excited states~\cite{peruzzo2014variational,higgott2018variational,o2018quantum}.   For a molecule with unknown electronic configuration, this is called the electronic structure problem~\cite{aspuru2005simulated,mcardle2018quantum} and it is crucially important in chemistry and material science.  However,  electronic structure Hamiltonians contain a very large number of terms and unfortunately the gate count of Trotter-Suzkui increases with the number of terms.  While the scaling is formally efficient, the required number of gates is impractically large.   An alternative to Trotter-Suzkui without this scaling problem would therefore have significant applications.   

A recurrent theme in the literature is that stochastic noise can be less harmful than coherent noise~\cite{wallman15,Knee15}, which hints that randomisation might be useful for washing out coherent errors in circuit design.  Poulin \textit{et al}~\cite{Poulin2011} showed that randomness is especially useful in simulation of time-dependent Hamiltonians as it allows us to average out rapid Hamiltonian fluctuations.  Campbell~\cite{campbell2017shorter} and Hastings~\cite{HastingMixing} have shown that random compiling can actually help reduce errors below what is feasible with a deterministic compiler. Since optimisation of Hamiltonian simulation circuits is a special case of compilation, one expects random compilers to be helpful in this setting. Following this line of reasoning, Childs, Ostrander and Su~\cite{childs2018faster} showed that it is useful to randomly permute the order of terms in Trotter-Suzuki decompositions.   However, randomly permuted Trotter-Suzuki decompositions still suffer the same scaling problem that plagues deterministic Trotter-Suzuki; that is, the gate count depends on the number of Hamiltonian terms.

Here we propose a simple and elegant approach to Hamiltonian simulation that uses randomisation to cure this scaling problem.  Our proposal is similar to Trotter-Suzuki in that we implement a sequence of small rotations, without any use of ancillary qubits or complex circuit gadgets.  Our key idea is to weight the probability of gates by the corresponding interaction strength in the Hamiltonian. Our simulation scheme can be seen as a Markovian process, which is inherently random but biased in such a way that we stochastically drift toward the correct unitary with high precision.  For this reason, we call it the quantum stochastic drift protocol, or simply qDRIFT.   Unlike any Trotter-Suzuki method, the gate count of qDRIFT is completely independent of the number of terms in the Hamiltonian.   Consequently, we find that our approach can speed-up quantum simulations of electronic structure Hamiltonians by several orders of magnitude within regimes of practical interest.  For example of the 60 qubit ethane, we find a speed-up of over a factor 1000 when the approximation error is $0.001$ and simulation time is  $t=6000$ (the same simulation time often used in phase estimation~\cite{wecker2014gate}). In quantum chemistry, phase estimation is performed using controlled $e^{i tH}$ unitaries and here our techniques can lead to even larger resource savings.

Our analysis is limited in scope in two ways.  First, we only compare against other Trotter-Suzuki decompositions. However, there are numerous approaches outside the Trotter-Suzuki family that make use of ancillary qubits and complex gadgets to obtain better asymptotic performance~\cite{berry2009black,berry2017exponential,berry2015simulating,berry2015hamiltonian,low2016hamiltonian,babbush2018encoding}, such as the LCU (linear combinations of unitary) technique.  Second, we only compare performance of rigorous bounds on gate counts, even though numerical studies of small systems show that far fewer gates are needed than suggested by rigorous bounds~\cite{poulin2014trotter,trotterError15,childs2018toward}.  Note that for the special case of local Hamiltonians, tighter analysis is possible because error propagation is localised and obeys Lieb-Robinson bounds~\cite{haah2018quantum,childs2019nearly}, but unfortunately electronic structure Hamiltonians are highly nonlocal.  

\begin{table}
	\begin{tabular}{|l|c|}		 \hline
		\textbf{Protocol} & \textbf{Gate count} (upper bound) 	 \\\hline
		1$^{st}$ order Trotter DET & $O(L^3 (\Lambda t)^2 / \epsilon)$  \\
		2$^{nd}$ order Trotter DET & $O(L^{5/2} (\Lambda t)^{3/2}  / \epsilon^{1/2} )$  \\
		(2k)$^{th}$ order Trotter DET & $O(L^{2 + \frac{1}{2k}} (\Lambda t)^{1 + \frac{1}{2k}}  / \epsilon^{1/2k} )$  \\  \hline
		(2k)$^{th}$ order Trotter RANDOM & $O(L^{2} (\Lambda t)^{1 + \frac{1}{2k}}  / \epsilon^{1/2k} )$  \\ \hline
		qDRIFT (general result) & $O( (\lambda t)^2 / \epsilon ) $  \\
		qDRIFT (when $\lambda=\Lambda L$) & $O(  L^2 (\Lambda t)^2 / \epsilon ) $  \\
		qDRIFT (when $\lambda= \Lambda \sqrt{L}$) & $O( L (\Lambda t)^2 / \epsilon ) $  \\ \hline
	\end{tabular}
	\caption{Resource scaling for different product formulae (see App.~\ref{App_Noise_measures} and ~\ref{App_Trotter_Suzuki} for details and caveats).}
\end{table}	

\textit{The Hamiltonian simulation problem}.- We begin by restating the problem more formally.  Consider a Hamiltonian
\begin{equation}
	H = \sum_{j=1}^L  h_j H_j 
\end{equation}
decomposed into a sum of $H_j$ each of which is Hermitian and normalised (such that the largest singular value of $H_j$ is 1).  We can always choose $H_j$ so that the weighting $h_j$ are positive real numbers.  Herein we denote $\lambda=\sum_j h_j$ and remark that this upper bounds the largest singular value of  $H$.  The decomposition of the Hamiltonian should be such that for each $H_j$ the unitary $e^{i \tau H_j }$ can be implemented on our quantum hardware for any $\tau$.  Our goal is then to find an approximation of $e^{ i t H }$ into a sequence of $e^{ i \tau H_j }$ gates up-to some desired precision.  We  use the number of $e^{ i \tau H_j }$ unitaries to quantify the cost of the quantum computation, and we aim to minimise the number of such unitaries used.  In the simplest Trotter formulae, one divides $U=e^{i  t H }$ into $r$ segments so that $U=U_r^r$ with $U_r=e^{ i t H/r }$ and uses that
\begin{equation}
		V_r = \prod_{j=1}^L e^{ i  t h_j H_j / r } ,
\end{equation}
approaches $U_r$ in the large $r$ limit.  Furthermore, $r$ repetitions of $V_r$ will approach $U$ in the large $r$ limit, so $V_r^r \rightarrow U$.  The gate count in this sequence will be $N=L r$, so we would like to know the smallest $r$ that suffices to achieve a desired precision $\epsilon$.  Analytic work on this problem (we use the analysis of Refs.~\cite{childs2018toward,childs2018faster}) shows that the Trotter error is no more than
\begin{equation}
   \epsilon = \frac{L^2 \Lambda^2 t^2}{2 r }  e^ {   \Lambda t L / r  } ,
\end{equation}
where $\Lambda := \mathrm{max}_j h_j$ is the magnitude of the strongest term in the Hamiltonian.  Solving for $r$ we find  approximately  $r \sim  L^2 \Lambda^2 t^2  / 2\epsilon $ segments are needed, each segments contains $L$ unitaries, leading to a total gate count of $N =Lr \sim L^3 (\Lambda t)^2  / 2 \epsilon$.   Table 1 compares this against other approaches including more sophisticated higher-order Suzuki decompositions.  As we increase the order of the decomposition, the scaling approaches $O(L^2 \Lambda t)$,  although the constant factors become rapidly worse for higher orders, so that in practice the optimal choice is usually second or fourth order.   Childs, Ostrander and Su,  showed that randomly permuted Trotter decompositions can further improve the gate count (see Table 1).

Having reviewed the prior art of product formaule, we notice the $L$ dependence never improved below quadratic.  Therefore, Trotter decompositions are limited to simulations of quantum systems with sparse interactions, so that $L$ must scale polynomially with the system size $n$. Furthermore, in chemistry problems $L=O(n^4)$ and while technically efficient, the resulting $O(n^8)$ scaling is prohibitively large.  Next we turn to our protocol that eliminates this dependence.

\begin{figure}[tb] \framebox{
		\begin{minipage}{.45\textwidth}
			\raggedright
			{\bf Input:} A list of Hamiltonian terms $H=\sum_j h_j H_j $, a classical oracle function SAMPLE() that returns an value $j$ from the probability distribution $p_j = h_j / (\sum_{j} h_j)$ and a target precision $\epsilon$.  \\
			{\bf Output:} An ordered list $V_{\mathrm{list}}$ of unitary gates of the form $\exp(i \tau H_j)$. \\
			\begin{enumerate}
				\item $\lambda \leftarrow \sum_j h_j$ 
				\item $N \leftarrow \lceil  2 \lambda ^2 t^2 / \epsilon  \rceil$   (or solve exact expression in appendix)
				\item $i \leftarrow 0$
				\item $V_{\mathrm{list}} = \{ \}$  (set gate list empty)
				\item While $i < N$
				\begin{enumerate}
					\item $i \leftarrow i+1$
					\item $j \leftarrow SAMPLE()$ 
					\item Append $e^{i \lambda t H_j / N }$ to ordered list $V_{\mathrm{list}}$
				\end{enumerate}	
				\item Return $V_{\mathrm{list}}$.
			\end{enumerate} 
	\end{minipage}  }
	\caption{Pseudocode for the qDRIFT protocol}
	\label{MyALGo}
\end{figure}

\begin{figure*} 
\includegraphics[width=\textwidth]{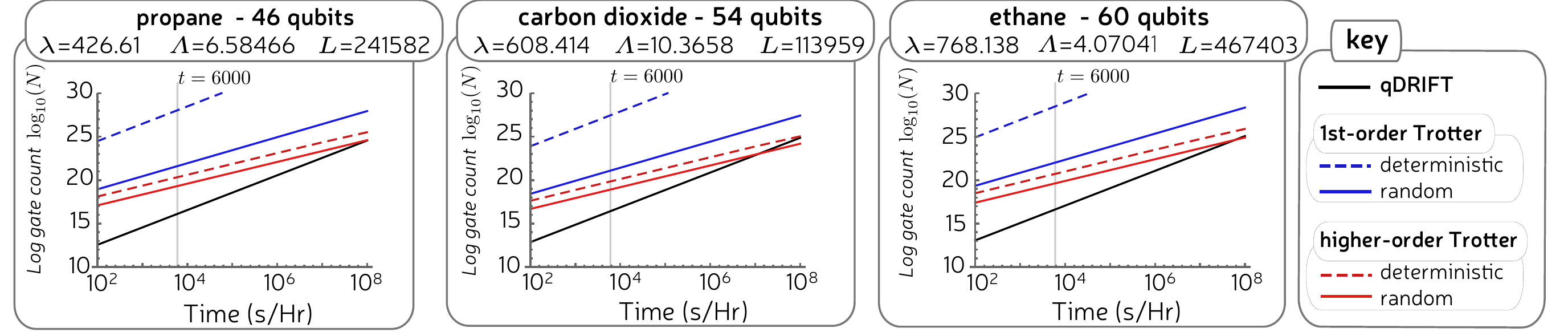}
	\caption{The number of gates used to implement $U=\exp(i H t)$ for various $t$ and $\epsilon=10^{-3}$ and three different Hamiltonians (energies in Hartree) corresponding to the electronic structure Hamiltonians of propane (in STO-3G basis), carbon dioxide (in 6-31g basis) and ethane (n 6-31g basis). Since the Hamiltonian contains some very small terms, one can argue that conventional Trotter-Suzuki methods would fare better if they truncate the Hamiltonian by eliminating negligible terms.  For this reason, whenever simulating to precision $\epsilon$ we also remove from the Hamiltonian the smallest terms with weight summing to $\epsilon$.  This makes a fairer comparison, though in practice we found it made no significant difference to performance. For the Suzuki decompositions we choose the best from the first four orders, which suffices to find the optimal.}
		\label{Fig_Molecule}
\end{figure*}

\textit{The qDRIFT protocol}.-  Our full algorithm is given as pseudocode in Fig.~\ref{MyALGo}.  Each unitary in the sequence is selected independently from an identical distribution (i.i.d sampling).   The strength $\tau_j$ of each unitary is fixed to a constant $\tau_j=\tau:=t \lambda /N$, which is independent of $h_j$ so, we implement gates of the form $e^{ i \tau   H_j }$.  The probability of choosing unitary  $e^{ i \tau   H_j }$ is weighted by the interaction strength $h_j$,  with normalisation of the distribution entailing that $p_j = h_j / \lambda$.  Therefore, the full circuit implemented is labelled by an ordered list of $j$ values $\vec{j}=\{ j_1 , j_2 , \ldots , j_N \}$ that corresponds to unitary
\begin{equation}
\label{Vj_unitary}
	 V_{\vec{j}} = \prod_{k=1}^N  e^{ i \tau   H_{j_k} }
\end{equation}	
which is selected from the product distribution $P_{\vec{j}} =\lambda^{-N} \prod_{k=1}^N  h_{j_k}$. While this quantum process is random, we build into the probabilities a bias so that with many repetitions the evolution stochastically drifts towards the target unitary.   Since each unitary is sampled independently, the process is entirely Markovian and we can consider the evolution resulting from a single random operation.  The evolution is mathematically represented by a quantum channel  that mixes unitaries as follows
\begin{align}
	 \mathcal{E}(\rho)  & = \sum_j p_j e^{  i \tau H_j } \rho   e^{ - i \tau H_j }  \\
	  & = \sum_j \frac{h_j}{ \lambda } e^{  i \tau H_j } \rho   e^{ - i \tau H_j }  .
\end{align}	
Using Taylor series expansions of the exponentials, we have that to leading order in $\tau$,
\begin{align}
\label{RandChannelRough}
\mathcal{E}(\rho)  & = \rho  +i  \sum_j \frac{h_j \tau}{ \lambda}   (H_j  \rho - \rho H_j ) + O(\tau^2) .
\end{align}	
We compare this with the channel $\mathcal{U}_N$ that is one $N^{\mathrm{th}}$ of the full dynamics we wish to simulate, so that
\begin{align}
   \mathcal{U}_N( \rho ) & =   e^{ i t H/N } \rho  e^{-i t H/N}  \\ \nonumber
    & = \rho +  i \frac{t}{N}( H \rho  - \rho H ) + O \left(  \frac{t^2 }{ N^2 } \right) ,
\end{align} 
where we have expanded out to leading order in $t/N$.  Using that $H= \sum_j h_j H_j$, we have
\begin{align}
\mathcal{U}_N( \rho )  & = \rho +  i  \sum_j  \frac{t h_j}{N} ( H_j \rho  - \rho H_j ) + O \left(  \frac{t^2 }{ N^2 } \right) .
\end{align} 
Comparing $\mathcal{E}$ and $\mathcal{U}_N$, we see that the zeroth and first order terms match whenever $\tau= t \lambda / N$.  The higher order terms will not typically match and more careful analysis (see App.~\ref{App_Noise_measures}) shows that the channels $\mathcal{E}$  and  $\mathcal{U}_N$ differ by an amount  bounded by
\begin{equation}
	\label{deltaBound}
\delta	 \leq  \frac{ 2 \lambda^2 t^2 }{ N^2 }  e^{ 2 \lambda t  / N } \approx  \frac{ 2 \lambda^2 t^2 }{ N^2 } ,
\end{equation}
where the first inequality is rigorous and the approximation on the right is very accurate  even for modest $N$.  

Since  $\delta$ is the approximation error on a single random operation $\mathcal{E}$,  the error of $N$ repetitions $\mathcal{E}^{N}$ relative to the target unitary $U$ is then
\begin{equation}
  \epsilon = N \delta \lesssim    \frac{ 2 \lambda^2 t^2}{N}  .
\end{equation}
We see the total error decreases as we increase $N$.   Setting $N$ to $N_{\mathrm{qD}} =  2 \lambda^2 t^2/\epsilon $ (rounding up to nearest integer) suffices to ensure that  $N \delta$ is less than the required precision $\epsilon$.  The exact value of $N$ is easily calculated, but again the aforementioned approximation is very good.


\textit{Asymptotics comparison}.- The qDRIFT approach needs approximately $2 \lambda^2 t^2   / \epsilon $ gates and we include this in Table 1 to compare against prior methods.  Since it does not explicitly depend on $L$, there are no sparsity constraints and this is the only known product formulae to beat the $O(L^2)$ barrier.  Though one may argue that $L$ dependence is hidden in $\lambda= \sum_j h_j$. The bounds for other Trotter-Suzuki formulae are given in terms of $\Lambda = \mathrm{max}_j h_j$, and these quantities are related by $ \lambda \leq \Lambda L$. The worst case for qDRIFT is therefore $\lambda=\Lambda L$, which occurs for systems like the 1D nearest neighbour Heisenberg chain~\cite{childs2018toward,childs2018faster,nam2018low}.  In this regime, qDRIFT is significantly better than first-order Trotter but the asymptotics suggest it will be outperformed by higher order Trotter. However, many real world systems have long range interactions that lead to $\lambda \ll \Lambda L$.   For instance, if we had $\lambda \sim \Lambda \sqrt{L}$ then the qDRIFT scaling would be $O(L)$, which is comfortably better than the $O(L^2)$ that was the best prior art.  While qDRIFT has significantly better $L$ dependence, it does depend quadratically on $\Lambda t$ whereas higher-order Trotter approaches linear scaling in $\Lambda t$.  Therefore, for a fixed Hamiltonian, qDRIFT may excel for short times, but there will always be a critical $t$ value above which it performs worse.  

 \textit{Numerics}.-  We have generated electronic structure Hamiltonians for propane, carbon-dioxide and ethane by using the openFermion library~\cite{mcclean2017openfermion}, which naturally satisfy $\lambda \ll \Lambda L$ and so qDRIFT should perform favourably.  We present our results in Fig.~\ref{Fig_Molecule} using target precision $\epsilon=10^{-3}$.  Observe that qDRIFT offers a significant advantage at low $t$, which is often several orders of magnitude better than any prior Trotter-Suzuki decomposition.  We remarked in our introduction that $t=6000$ has been identified as relevant for phase estimation in quantum chemistry problems~\cite{wecker2014gate} and here we see speed-ups of $591 \times $, $306\times $ and $1006\times $  for propane, carbon dioxide and ethane (respectively).    However, since qDRIFT scales worse with $t$ than higher order Trotter, for  longer time simulations our advantage decreases and we eventually observe a cross-over at times around $t=10^7-10^8$ where prior methods perform better.  But this cross-over does not occur until the simulation time is so long that $10^{23}-10^{25}$ gates are required.  This is an extremely high gate count.  Quantum error correction would certainly be needed and it is well known that to implement this many non-Clifford gates would require many billions of physical qubits even with generous hardware assumptions~\cite{Fowler12,Ogorman16,reiher2017elucidating,campbell2018applying}.  For these molecules, any foreseeable device performing Hamiltonian simulation would significantly benefit from using qDRIFT over standard Trotter-Suzuki. 
 
 \textit{Phase estimation}.- When using phase estimation to find ground state energies, one performs many controlled-$\exp(iHt)$ rotations.  Estimating energies to precision $\delta_E$ --- chemical precision means $\delta_E \sim 10^{-4}$ ---  the largest time used is at least $t \sim \pi / \delta_E$, with slightly longer times needed to boost the inherent success probability of phase estimation.  Note that the Trotter error $\epsilon$ is not directly connected to $\delta_E$ but instead contributes to the failure probability.  Running phase estimation several times allows us to handle modest failure probabilities, so in practice $\epsilon$ can be much larger than $\delta_E$.  Therefore, the relevant $\epsilon$ and $t$ regime for phase estimation matches the regime where qDRIFT performs well in simulation tasks.   We provide a detailed analysis of phase estimation in App.~\ref{App_Noise_measures}, which shows that qDIRIFT offers $2-3$ orders of magnitude improvement when the failure probability of a single run is 5\%.
 
\textit{Diamond norm distance.-} An important technicality is that for a random circuit the appropriate measure of error $\epsilon$ is the diamond norm distance~\cite{watrous2018theory}.  If we instead consider a specific instance of a randomly chosen unitary $V_{\vec{j}}$ in Eq.~\eqref{Vj_unitary}, then the error will typically (on average) be much larger than $\epsilon$, with standard statistical arguments (see e.g.~\cite{Poulin2011}) suggesting it would be closer to $\sqrt{\epsilon}$.  It is counter-intuitive that the random circuit error is considerably less than the error of any particular unitary, so let us elaborate.  If we initialise the quantum computer in state $\ket{\psi}$, then qDRIFT leads to state $\ket{\Psi_{\vec{j}}}=V_{\vec{j}}\ket{\psi}$ with probability $P_{\vec{j}}$.  If our experimental setup forgets (erases from memory) which unitary was implemented, then it prepares the mixed state
 \begin{equation}
 	\rho =\mathcal{E}^{N} ( \kb{\psi}{\psi} ) = \sum_{\vec{j}} P_{\vec{j}} V_{\vec{j}} \kb{ \psi  }{ \psi } V_{\vec{j}}^\dagger = \sum_{\vec{j}} P_{\vec{j}} \kb{ \Psi_{\vec{j} } }{ \Psi_{\vec{j} } } .
 \end{equation}	 
 Since this channel is $\epsilon$-close in diamond distance to the ideal channel $\mathcal{U}$, it follows that $\rho$ is $\epsilon$-close in trace norm distance to the target state $\mathcal{U}(\kb{\psi}{\psi})=U\kb{\psi}{\psi}U^\dagger$.   Trace norm distance is the relevant quantity because it ensures that if we perform a measurement, then the probabilities of the outcomes (on state $\rho$) do not differ by more than $2 \epsilon$ from the ideal probability given by $U\ket{\psi}$.  Provided we estimate expectation values over several runs, each using a new and independent randomly generated unitary, the precision of our estimate will be governed by $\epsilon$ rather than the looser $\sqrt{\epsilon}$ bound obtained without use of the diamond norm.
 
\textit{Discussion.-}    A common setting is where $H_j$  are taken as tensor products of Pauli spin operators, then $e^{ i \tau H_j }$ can be realised using Clifford gates and a single-qubit Pauli $Z$ rotation~\cite{RS14}.  When performing quantum error correction, the resource overhead of Clifford gates is negligible~\cite{Fowler12,Ogorman16} whereas the single-qubit Pauli $Z$ rotation must be decomposed into a large number of single-qubit $T$ and Clifford gates.   One further advantage of qDRIFT is that it consumes many Pauli rotations of exactly the same angle,  allowing the use of adder-circuit catalysis that significantly reduce $T$-counts~\cite{gidney2017halving,beverland2019lower}. This is especially true when the Pauli rotations then belong to the Clifford hierarchy~\cite{CliffHier}, since one then has the option of directly distilling magic states providing the rotation without further compilation~\cite{landahl13,duclos15,campbell16,campbell2018magic}.  Interestingly, Duclos-Cianci and Poulin~\cite{duclos15} give a short discussion of how their magic state distillation protocol could be used in a Hamiltonian simulation scheme using a modified-Trotter decomposition where the gates all have the same $\tau$ value.   While they allude to such a Hamiltonian simulation protocol, they do not provide any details or error analysis and nor did they suggest that randomisation would be part of the protocol. 

\textit{Acknowledgements}.- This work was supported by the EPSRC (grant no. EP/M024261/1).  We thank Simon Benjamin, Xiao Yuan and Sam McArdle, for discussions on the electronic structure problem and providing molecular Hamiltonians taken from openFermion.  For regular discussions on Hamiltonian simulation we thank John Clark, David White, Ben Jones and George O'Brien. We thank Yuan Su for sharing details regarding Ref.~\cite{childs2018faster}.   For comments on the manuscript, we thank Dominic Berry.
 
 
 %

\appendix

\section{Error measures}	
\label{Error_measures}

Here we switch to more mathematical notation than used in the main text.  We use $|| \ldots ||$ to denote the operator norm or Schatten-$\infty$ norm, which is equal to the largest singular value of an operator.   We use $|| \ldots ||_1$ for the trace norm or Schatten 1-norm,   defined as $|| Y ||_1 := \mathrm{Tr}[  \sqrt{ Y^\dagger  Y } ]$, which is equal to the sum of the singular values of an operator.  Throughout, we use the diamond norm distance as a measure of error between two channels.  The diamond distance is denoted
\begin{equation}
d_{\diamond}( \mathcal{E} ,  \mathcal{N} ) = \frac{1}{2} || \mathcal{E}- \mathcal{N}  ||_\diamond ,
\end{equation}
where $|| \ldots ||_\diamond$ is the diamond norm
\begin{equation}
|| \mathcal{P} ||_\diamond := \mathrm{sup}_{\rho ; ||\rho||_1=1 }   || ( \mathcal{P} \otimes \id )( \rho )||_1 ,
\end{equation}
where $\id$ acts on the same size Hilbert space as $\mathcal{P}$.    We are using curly script such as $\mathcal{P}$ to denote superoperators, and will use $\mathcal{P}^n$ to denote $n$ repeated applications of a superoperator. Two key properties of the diamond norm that we employ are:
\begin{enumerate}
	\item The triangle inequality: $|| \mathcal{A} \pm  \mathcal{B} ||_{\diamond} \leq || \mathcal{A} ||_{\diamond}+ || \mathcal{B} ||_{\diamond}$,
	\item Sub-multiplicativity: $|| \mathcal{A}  \mathcal{B} ||_{\diamond} \leq || \mathcal{A}  ||_{\diamond} || \mathcal{B}  ||_{\diamond}$ and consequently  $ || \mathcal{A}^n  ||_{\diamond} \leq  || \mathcal{A}  ||_{\diamond}^n$.
\end{enumerate}	
From the definition of diamond distance it follows that if we apply the channels $\mathcal{E}$ and $\mathcal{N}$ to quantum state $\sigma$, we have that 
\begin{equation}
d_{\mathrm{tr}}( \mathcal{E}(\sigma) , \mathcal{N}(\sigma)  ) = \frac{1}{2} ||  \mathcal{E}(\sigma)  - \mathcal{N}(\sigma)  ||_1 \leq  d_{\diamond}( \mathcal{E} ,  \mathcal{N} ) .
\end{equation}		
The trace norm distance is an important quantity because it bounds the error in expectation values.  If $M$ is an operator, then
\begin{align}
	| \mathrm{Tr} [  M \mathcal{E}(\sigma)  ] - \mathrm{Tr} [  M \mathcal{N}(\sigma)  ] |  & \leq  2 || M || d_{\mathrm{tr}}( \mathcal{E}(\sigma) , \mathcal{N}(\sigma)  ) \\ \nonumber
	& \leq  2 || M || d_{\diamond}( \mathcal{E} ,  \mathcal{N} ) .
\end{align}	
If $M$ is a projection so that this represents a probability, then $||M||=1$.  We see $\epsilon$ error in diamond distance ensures that the measurement statistics are correct upto additive error $2 \epsilon$.

\section{Bounding higher order error terms}
\label{App_Noise_measures}

Next, we make use of the Liouvillian representation of a unitary channel so that
\begin{equation}
e^{i H t} \rho e^{-i H t} = e^{t \mathcal{L}}(\rho) = \sum_{n=0}^\infty \frac{t^n \mathcal{L}^n(\rho)}{n!}  ,
\end{equation}	
where
\begin{equation}
\mathcal{L}(\rho)= i (H\rho - \rho H) .
\end{equation}	
We have that 
\begin{equation} 
\label{BigLNORM}
|| \mathcal{L} ||_\diamond \leq 2 || H || \leq 2 \lambda .
\end{equation}	
Similarly, we can define $\mathcal{L}_j$ that generate unitaries under Hamiltonians $H_j$ so that 
\begin{equation}
\label{Eq_SumLj}
\mathcal{L} = \sum_j h_j \mathcal{L}_j
\end{equation}	
and 
\begin{equation} 
\label{LjNORM}
|| \mathcal{L}_j ||_\diamond \leq 2 || H_j || \leq 2 .
\end{equation}

We will now upperbound the error of the qDRIFT protocol, though remark that a very similar upperbound can be found by employing the Hastings-Campbell mixing lemma~\cite{campbell2017shorter,HastingMixing}.   Each random operator of qDRIFT implements a single randomly chosen gate so that 
\begin{equation}
\mathcal{E} = \sum_j p_j e^{\tau  \mathcal{L}_j  } =  \sum_j \frac{h_j}{\lambda} e^{\tau  \mathcal{L}_j  } ,
\end{equation}	
which expands out to
\begin{align}
	\mathcal{E} & = \id + \left(\sum_j \frac{h_j \tau}{\lambda} \mathcal{L}_j  \right) + \sum_j \frac{h_j}{\lambda} \sum_{n=2}^\infty \frac{\tau^n \mathcal{L}_j^n}{n!}     \\
	& =  \id + \frac{\tau}{\lambda} \mathcal{L} + \sum_j \frac{h_j}{\lambda} \sum_{n=2}^\infty \frac{\tau^n \mathcal{L}_j^n}{n!}  ,
\end{align}	
where in the second line we have used Eq.~\eqref{Eq_SumLj}.
This is to be compared against 
\begin{align}
	\mathcal{U}_N & = e^{t \mathcal{L}/N } \\ \nonumber
	&=  \id + \frac{t}{N}  \mathcal{L} + \sum_{n=2}^\infty   \frac{t^n \mathcal{L}^n }{n! N^n } 
\end{align}	
We see the first two terms of $\mathcal{E}$ and  $\mathcal{U}_N$ will match whenever $\tau= \lambda t / N$.  Using this value for $\tau$, we have
\begin{align}  \nonumber
	|| \mathcal{U}_N - 	\mathcal{E} ||_\diamond & = \bigg| \bigg|  \sum_{n=2}^\infty \frac{t^n \mathcal{L}^n }{n! N^n }  -  \sum_j \frac{h_j}{\lambda} \sum_{n=2}^\infty \frac{\lambda^n t^n \mathcal{L}_j^n}{n! N^n}   \bigg| \bigg|_\diamond \\ \nonumber
	&	\leq  \sum_{n=2}^\infty  \frac{t^n ||\mathcal{L}^n ||_\diamond}{ n! N^n }  +  \sum_j \frac{h_j}{\lambda} \sum_{n=2}^\infty \frac{\lambda^n t^n ||\mathcal{L}_j^n||_\diamond}{n! N^n}  
\end{align}
The first inequality uses the triangle inequality and that all variables are positive real numbers. Next we use sub-multiplicativity combined with Eq.~\eqref{BigLNORM} and Eq.~\eqref{LjNORM} to conclude that  $||\mathcal{L}^n ||_\diamond \leq ||\mathcal{L} ||_\diamond^n \leq  (2 \lambda)^n$ and $||\mathcal{L}_j^n ||_\diamond \leq ||\mathcal{L}_j ||_\diamond^n \leq 2^n$, which leads to 
\begin{align}  \nonumber
	|| \mathcal{U}_N - 	\mathcal{E} ||_\diamond &	\leq  \sum_{n=2}^\infty \frac{1}{n!} \left( \frac{ 2 \lambda t }{ N } \right)^n +  \sum_j \frac{h_j}{\lambda} \sum_{n=2}^\infty \frac{1}{n!} \left( \frac{2 \lambda t }{N}  \right)^n \\ \nonumber
	& = 2  \sum_{n=2}^\infty \frac{1}{n!} \left( \frac{ 2 \lambda t }{ N } \right)^n.
\end{align}
The last equality uses that $\sum_j h_j =\lambda$ and collects together the pair of equal summations.  Since our definition of diamond distance includes a factor 1/2, we have
\begin{equation}
d(\mathcal{U}_N , 	\mathcal{E} ) \leq  \sum_{n=2}^\infty \frac{1}{n!} \left( \frac{ 2 \lambda t }{ N } \right)^n .
\end{equation}	
Next, we use the exponential tail bound (see Lemma F.2 of Ref~\cite{childs2018toward}) that states that for all positive $x$ we have
\begin{equation}
\sum_{n=2}^\infty \frac{x^n}{n!}	 \leq  \frac{x^2}{2}e^{x} ,
\end{equation}	
which we use with $x=2 \lambda t / N$ so that
\begin{equation}
d(\mathcal{U}_N , 	\mathcal{E} ) \leq   \frac{ 2 \lambda^2 t^2 }{ N^2 }  e^{ 2 \lambda t  / N } \approx  \frac{ 2 \lambda^2 t^2 }{ N^2 } 	.
\end{equation}
The approximation on the right is very accurate in the large $N$ limit.  This gives the result stated in the main text. Since the diamond distance is subadditive~\cite{watrous2018theory} under composition we have that
\begin{align}
	d_{\diamond} (\mathcal{U} , \mathcal{E}^N )   & \leq N 	d(\mathcal{U}_N , 	\mathcal{E} )  \\ \nonumber
	&=  \frac{ 2 \lambda^2 t^2 }{ N }  e^{ 2 \lambda t  / N } \approx  \frac{ 2 \lambda^2 t^2 }{ N }.
\end{align}

\section{Bounding higher order error terms}
\label{App_Trotter_Suzuki}

Here we reproduce for convenience some results on the Trotter and Suzuki decompositions.   All these results are taken from  Childs, Ostrander and Su~\cite{childs2018faster}.

First, we consider the Trotter decomposition, and begin by defining
\begin{align}
	a_{\mathrm{TROTT}} & := \frac{(L \Lambda t)^2}{r^2} e^{ \Lambda t / r  } , \\ \nonumber
	b_{\mathrm{TROTT}} & := \frac{(L \Lambda t)^3}{3 r^3} e^{    \Lambda t / r  }  .
\end{align}	
From this, one can show that deterministic and randomised Trotter decompositions have errors 
\begin{align}
	\epsilon_{\mathrm{TROTT}}^{\mathrm{det}} & \leq \frac{r}{2}a_{\mathrm{TROTT}}  , \\ \nonumber
	\epsilon_{\mathrm{TROTT}}^{\mathrm{random}} & \leq \frac{r}{2}(a_{\mathrm{TROTT}}^2 + 2  b_{\mathrm{TROTT}}).
\end{align}
One can see that if $b_{\mathrm{TROTT}} \ll a_{\mathrm{TROTT}} $ there is a significant advantage to the randomised approach.  To determine gate counts one must solve to find the smallest integer $r$ such that the errors are below some target $\epsilon$.  Since the Trotter decomposition has $r$ segments and each segment contains $L$ gates, the total gate count is $Lr$.

Next, we consider the $2k$-order Suzuki decompositions, starting with the definitions
\begin{align}
	a_{2k-\mathrm{SUZUKI}} & := 2 \frac{(2 \cdot 5^{k-1} (\Lambda t) L)^{2k+1}}{(2k+1)! (r^{2k+1}) }  e^{ 2 \cdot 5^{k-1} \Lambda t / r }   
	, \\ \nonumber
	b_{2k-\mathrm{SUZUKI}} & :=  \frac{(2 \cdot 5^{k-1} (\Lambda t) )^{2k+1} L^{2k}}{(2k-1)! (r^{2k+1}) } e^{ 2 \cdot 5^{k-1} \Lambda t / r }   .
\end{align}	
From this, one can show that deterministic and randomised $2k$-order Suzuki decompositions have errors bounded by
\begin{align}
	\epsilon_{2k-\mathrm{SUZUKI}}^{\mathrm{det}} & \leq \frac{r}{2}a_{2k-\mathrm{SUZUKI}}  , \\ \nonumber
	\epsilon_{2k-\mathrm{SUZUKI}}^{\mathrm{random}} & \leq \frac{r}{2}(a_{2k-\mathrm{SUZUKI}}^2 + 2  b_{2k-\mathrm{SUZUKI}}).
\end{align}
Again,  if $b_{2k-\mathrm{SUZUKI}} \ll a_{2k-\mathrm{SUZUKI}} $ there is a significant advantage to the randomised approach. However, in the limit $k \rightarrow \infty$ both $b_{2k-\mathrm{SUZUKI}}$ and $a_{2k-\mathrm{SUZUKI}} $ approach a similar order of magnitude.  As such, the advantage of randomised Suzuki decompositions disappears as $k$ increases, which was numerically reported by Childs, Ostrander and Su~\cite{childs2018faster}.   The other salient point is how the two terms of $\epsilon_{2k-\mathrm{SUZUKI}}^{\mathrm{random}}$ compare in size.  In different limits of $\Lambda, L$ and $\epsilon$, either the first or second term can dominate.  For the  $\Lambda$ and  $L$ set by the chemistry problems in the main text and with $\epsilon < 10^{-2}$, we find that the error is dominated by the second term, so a good approximation is given by 
\begin{align}
	\epsilon_{2k-\mathrm{SUZUKI}}^{\mathrm{random}} & \lessapprox r  b_{2k-\mathrm{SUZUKI}} , \\ 
	& \approx B_k  \frac{(\Lambda t) ^{2k+1} L^{2k}}{r^{2k} } ,
\end{align}
where in the second line we have also neglected the exponential (valid when $ \Lambda t \ll r$) and collected the constants into
\begin{align}
	B_k	& = \frac{( 2 \cdot 5^{k-1})^{2k+1}}{(2k-1)!} .
\end{align}
For chemistry problems we find this approximation to be very close to the exact upper bound.  We reiterate that the numerics presented in the main text used the exact expressions, but to gain intuition and study phase estimation these approximations are very useful.

To determine gate counts one must solve to find the smallest integer $r$ such that the errors are below some target $\epsilon$.  Using the above approximation one obtains
\begin{align}
	r	& = \Lambda t L \left( \frac{\Lambda t   B_k}{\epsilon} \right)^{\frac{1}{2k}} ,
\end{align}
where herein we drop the subscripts on $\epsilon$.  Since the $2k$-order Suzuki decompositions have $r$ segments and each segment contains 
$2 \cdot 5^{k-1 }L$ gates, the total gate count is $2 \cdot 5^{k-1 }Lr$, so we obtain a gate count
\begin{align}
	\label{NkHigher}
	N_k	& =  2 \cdot 5^{k-1 } \Lambda t L^2 \left( \frac{\Lambda t   B_k}{\epsilon} \right)^{\frac{1}{2k}} \\ 
	& =  C_k \frac{ L^2   (\Lambda t)^{1 + \frac{1}{2k}} }{\epsilon^{\frac{1}{2k}}}
\end{align}
where $C_k$ is the new constant $C_k= 2 \cdot 5^{k-1 }  B_k^{1/2k}$.   For instance we have
\begin{align} \label{eqN1}
	N_1 & = \frac{4 \sqrt{2} (\Lambda t)^{3/2} L^2}{\sqrt{\epsilon }} , \\
	N_2 & = \frac{500 \sqrt[4]{10} (\Lambda t)^{5/4} L^2}{3 \	\sqrt[4]{\epsilon }} ,\\
	N_3 & = \frac{156250 \sqrt[6]{2} \sqrt[3]{5} (\Lambda t)^{7/6} L^2}{3  \sqrt[6]{\epsilon }} .
\end{align}	
The scaling with $\Lambda, t$ and $\epsilon$ improves with $k$, but the constant prefactor becomes large, so in practice one rarely wishes go above $k=3$ and for modest $t$ and $\epsilon^{-1}$ values the optimal is often just the $k=1$ protocol.

\begin{figure} 
	\includegraphics{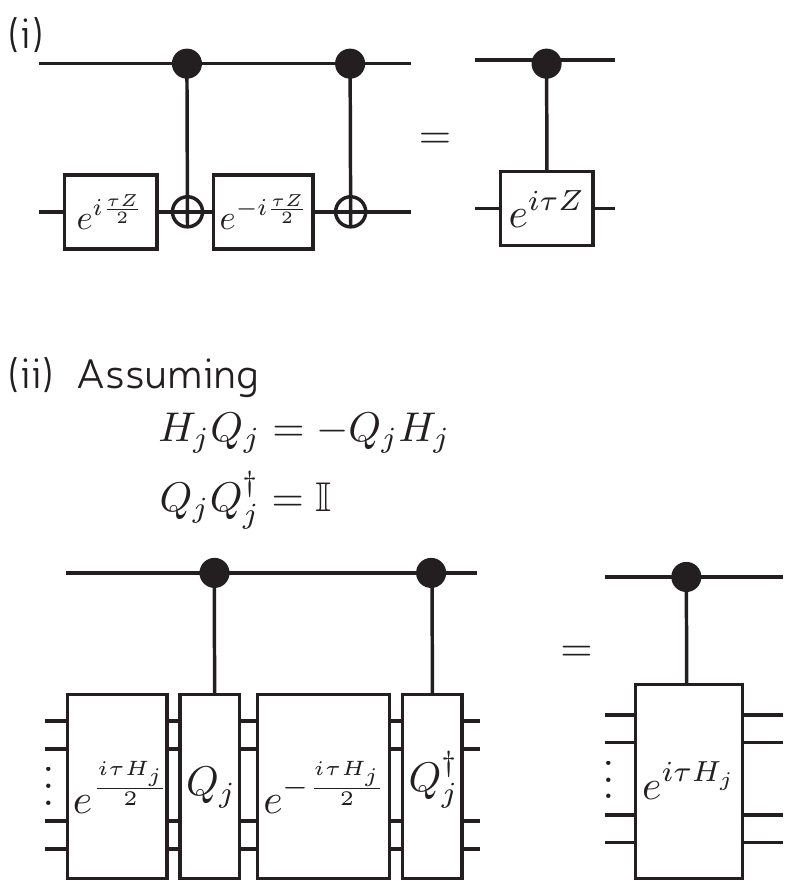}
	\caption{Implementing controlled rotations used in phase estimation.  (i) a simple circuit for implementing a controlled-$\exp(i \tau Z)$ gate using two single qubit $Z$ rotations and two control-X gates.  (ii) A more general circuit for implementing controlled-$\exp(i \tau H_j)$, assuming the existence of a suitable $Q_j$ operator and the ability to perform control-$Q_j$ and control-$Q_j^\dagger$.  Typically, we decompose our Hamiltonian into $H_j$ Pauli operators, in which case $Q_j$ and $Q_j^\dagger$ can be taken to be single qubit $X$ or $Z$ operators.  Therefore, the decomposition will use two $\exp(\pm i \tau H_j / 2)$ rotations and two control-X (or control-Z) gates.}
	\label{Fig_PhaseEst}
\end{figure}

\section{Controlled evolution}
\label{Sec_control}

To perform phase estimation we need to implement a controlled-$\exp(i H t)$ gate, but our analysis has shown only how to approximate $\exp(i H t)$ using qDRIFT.   We first observe that controlled-$\exp(i H t)$ is equal to $\exp(i (\kb{1}{1}  \otimes H) t)$.  Therefore, we can perform phase estimation by using qDRIFT with the Hamiltonian 
\begin{align}
	H' & = \kb{1}{1}  \otimes H \\ \nonumber
	& = \kb{1}{1}  \otimes (\sum_j h_j H_j) \\ \nonumber
	& = \sum_j h_j  H'_j 
\end{align}	
where  $H'_j= \kb{1}{1}  \otimes  H_j$.
Note that $||H_j||=1$ was already assumed and implies that $||H'_j||=1$.  Furthermore, $\lambda=\sum_j |h_j|$ and $L$ are unchanged.   This allows us to decompose the phase estimation circuit into a random product of exponentials $\exp(i \tau H'_j)= \exp(i (\kb{1}{1}  \otimes H_j) \tau )$.  Therefore, we see that for a given $t$ and $\epsilon$, controlled evolution needs exactly the same number of $\exp(i H_j' \tau )$ rotations as the number of $\exp(i H_j \tau )$ rotations as were needed for simulation.  However, perhaps our hardware can not natively implement $\exp(i H'_j \tau )$, in which case there is some additional overhead. However, $H_j$ are usually Pauli operators, in which case this can be achieved as in Fig.~\ref{Fig_PhaseEst} with constant factor overhead.

\section{ Phase estimation}
\label{App_PhaseEst}

Here we analyse and compare using qDRIFT and $2^{\mathrm{nd}}$-order random Trotter to implement a simple version of phase estimation in order to perform ground state estimation.  We follow the phase estimation protocol and borrow results from Cleve \textit{et. al.}~\cite{cleve1998quantum}, though we assume that classical feedforward is used instead of performing the quantum fourier transform (see Fig. 1c of Ref.~\cite{higgins2007entanglement}).  There have been many subsequent variants of phase estimation proposed that could significantly reduce the resource overhead, but our purpose here is just to demonstrate the utility of qDRIFT rather than give a detailed literature survey of phase estimation techniques.

When using phase estimation to solve the electronic structure problem for Hamiltonian $H$, we wish to find the energy $E_0$ of the ground state $\ket{\psi_0}$.  We do not know $\ket{\psi_0}$ but can prepare an ansatz state $\ket{\psi} = \sum_j c_j \ket{\psi_j}$ that has high overlap with the groundstate, so $f=|c_0|^2 \gg 0$.  Phase estimation aims to sample from the energies $E_j$ with some probability close to $|c_j|^2$.  Roughly, the idea is to perform phase estimation several times and take the lowest reported energy.  However, we can only estimate $E_j$ to finite precision and there is always some probability of failure.  It is useful to define  $A:=(H/\lambda + \id)/2$, which has eigenvalues in the range 0 to 1.   Both $H$ and $A$ share the same eigenstates, and estimating eigenvalues of $A$ to additive error $\delta$ enables us to estimate the energies to additive error $\delta_E=2 \lambda \delta $ where we typically want $\delta_E \leq 10^{-4}$ for chemical accuracy.  Given our target $\delta$ we translate this into a number of bits of precision $n = \log_2(\delta) -1$, rounded up. The more bits $n$, the more gates are needed in the phase estimation procedure.  However, phase estimation also has some inbuilt failure probability $p_f$ that can be suppressed by using a deeper algorithm.  Following Cleve \textit{et. al.}, the depth of the algorithm is determined by
\begin{align}
	\label{EquationM}
		m & = n + \log_2 \left( \frac{1}{2p_f} + \frac{1}{2}  \right)  \\
		& =   \log_2(\delta^{-1}) -1 + \log_2 \left( \frac{1}{2p_f} + \frac{1}{2}  \right) \\
		& =   \log_2(\delta^{-1}) + \log_2 \left( \frac{1}{p_f} + 1  \right) - 2
\end{align}	
rounded up.   The phase estimation protocol uses a sequence of control-$U^{2^{j-1}}$ unitaries where $U=\exp(i 2 \pi A)$ and $j=1,\ldots m$.  We will also write $U^{2^{j-1}}= \exp(i  A t_j)$ where $t_j=2^{j} \pi$.

The above discussion assumes no Trotter error.  Finite Trotter error can increase the probability of measuring incorrect outcomes.   Using the diamond norm bounds given earlier,  the total failure probability is bounded by 
\begin{align}
	\label{PFequation}
	P_f = p_f + 2  \epsilon_{\mathrm{tot}} =  p_f + 2 \sum_j \epsilon_j ,
\end{align}	
where $ \epsilon_{\mathrm{tot}} $ is the total Trotter error summed over all the control unitaries and $\epsilon_j$ is the Trotter error for control-$U^{2^{j-1}}$.

\subsection{Failure probabilities}
\label{SecFailureProbs}

The value of $P_f$ can be quite large without undermining the ground state estimation procedure and we give a rough overview of the statistics involved.  As remarked above, we will repeat phase estimation many times.  We would need to perform it at least $1/f$ times to be confident that we have sampled the ground state energy.  Given a finite failure probability, we need to repeat more times.  For instance, we could perform the following procedure: repeat phase estimation $M$ times and record the frequency $\nu(E)$ that we observe outcome $E$; output the smallest observed $E$ such that $\nu(E)>P_f+1/M$.   The $\nu(E)>P_f+1/M$ rule will filter out false energies.  The expected frequency of measuring the ground state energy satisfies $\nu(E_0) \geq f - P_f$, provided $f > 2 P_f + 1/M$ this approach will (with high probability) ensure that $\nu(E_0)>P_f+ 1/M$ and so the ground state energy will not be filtered out.   It is believed that single-determinant Hartree-Fock or known multi-determinant ansatz states usually achieve $f > 1/2$~\cite{tubman2018postponing} so $P_f$ can be quite large (e.g. $P_f \sim 5\%-10\%$) compared to $\delta_E$.

\subsection{qDRIFT}

For each control-$U^{2^{j-1}}$ unitary, let $N(j)$ denote the number of require gates to achieve the desired $\epsilon_j$.  For qDRIFT, we have
\begin{align}
			N(j) & = 2 \frac{ 2 \lambda_A^2 t_j^2 }{ \epsilon_j } =  \frac{   (2^{j} \pi  )^2 }{ \epsilon_j },  \\ \nonumber
  & =  \frac{   4^{j} \pi^2   }{ \epsilon_j } .
\end{align}	
Here the extra factor of 2 comes from Fig.~\ref{Fig_PhaseEst}.  We note that the relevant $\lambda$ is that of the operator $A$  -- ignoring the identity component --- and so $\lambda_A=1/2$.  We also use $t_j=\pi 2^{j}$.  We wish to select $\epsilon_j$ that minimizes $\sum_j N(j)$ subject to the constraint $\sum_j \epsilon_j = \epsilon_{\mathrm{tot}}$ and it is easy to confirm that this is achieved by setting
\begin{equation}
	\epsilon_j = \epsilon_{\mathrm{tot}} \frac{2^j}{2(2^m -1)} .
\end{equation}	
This leads to
\begin{align}
	N(j) &  =2 \frac{   2^{j} \pi^2  (2^m - 1) }{ \epsilon_{\mathrm{tot}} } .
\end{align}	
Summing over all $j$ from 1 to $m$, we get
\begin{align}
N = \sum_{j=1}^{m}N(j) & = 4 \frac{   \pi^2 (2^m - 1)^2 }{  \epsilon_{\mathrm{tot}} } 
\end{align}	
Using Eq.~\eqref{EquationM} to substitute in a value for $m$, we find $2^m$ is
\begin{align} 	\label{Eq2m}
	 2^m  & = \frac{1}{4\delta}\left( \frac{1}{ p_f} + 1  \right) \\
	  & = \frac{\lambda}{2 \delta_E}\left( \frac{1+p_f}{ p_f}  \right) ,
\end{align}	
Since $\delta_E \leq 10^{-4}$ for chemical accuracy, we have that $2^m \gg 1$  and we can take $2^m - 1 \sim 2^m$. Therefore,
\begin{align}
	N & = \frac{   \pi^2  \lambda^2 }{  \epsilon_{\mathrm{tot}} \delta_E^2 }  \left( \frac{1+p_f}{ p_f}  \right)^2 \\ \nonumber
	& = \frac{   \pi^2  \lambda^2 }{  \delta_E^2 }  \left( \frac{1+p_f}{ p_f  \sqrt{\epsilon_{\mathrm{tot}}}}  \right)^2
\end{align}	
Let us define the term in the large brackets as
\begin{align}
	X := \frac{1+p_f}{ p_f  \sqrt{\epsilon_{\mathrm{tot}}}} .
\end{align}	
Using Eq.~\eqref{PFequation} to eliminate $\epsilon_{\mathrm{tot}}$ in favour of $P_f$ we have
\begin{equation}
X =	\frac{1+p_f}{p_f \sqrt{\epsilon_{\mathrm{tot}}}}  = \sqrt{2} \frac{1+p_f}{p_f \sqrt{P_f - p_f}}   ,
\end{equation} 
We want to minimise $X$ over all $0 \leq p_f < P_f$ and treating $P_f$  as a constant.   The exact minimal value of $X$ is involved, but assuming small $P_f$ the optimal solution is given by $p_f=(2/3)P_f$.   Then the minimal solution satisfies $X^2 \leq 27  / 2P_f^{3}$ in the small $P_f $ regime and this is fairly accurate for modest size $P_f$.  Putting this together yields
\begin{align}
	\label{qDRIFTphase}	N & \sim    \frac{ 27  \pi^2 }{ 2  }   \frac{  \lambda^2 }{  \delta_E^2 P_f^3 }   \\ \nonumber
 &	\sim 133   \frac{  \lambda^2 }{  \delta_E^2 P_f^3 }  ,
\end{align}	
where in the last line we have collected the constants and rounded to the first three significant figures.

\begin{figure*} 
	\includegraphics[width=\textwidth]{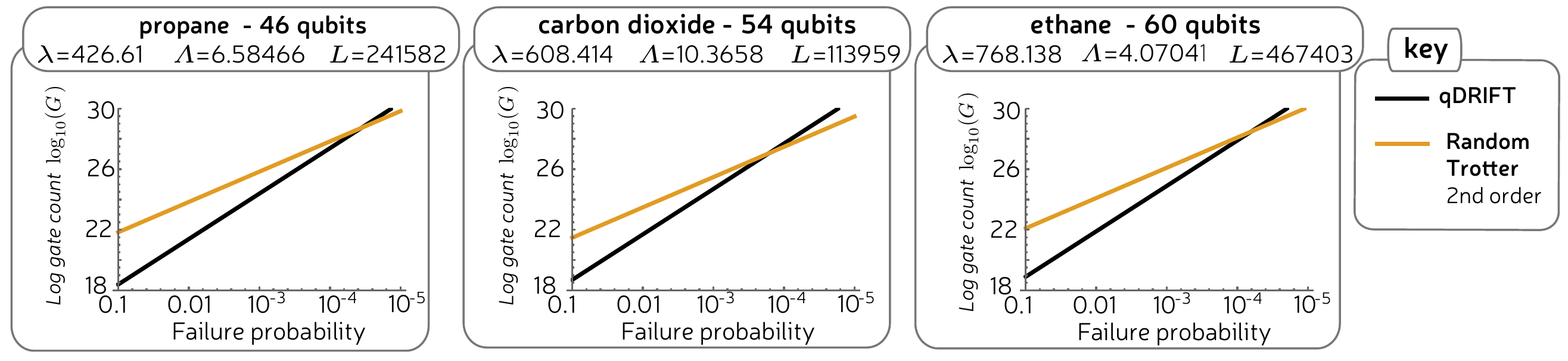}
	\caption{The number of gates used to perform phase estimation with $\delta_E=10^{-4}$ as a function of the failure probability.  }
	\label{Fig_PhaseCosts}
\end{figure*}

\subsection{Random Trotter}

Next, we follow the same analysis as in the previous section but for second order random Trotter.  Then the gate count for control-$U^{2^{j-1}}$ gate is bounded by
\begin{align}
	N(j) & =2 \cdot  4 L^2  \left( \frac{2  \Lambda_A^3 t_j^3 }{\epsilon_j}  \right)^{\frac{1}{2}}
\end{align}	
where the first factor 2 again comes from Fig.~\ref{Fig_PhaseEst} and the rest of the expressed is given by Eq.~\eqref{eqN1}. Here $\Lambda_A$ is for the renormalised $H$ and so 
\begin{equation}
	\label{EqlambdaA}
	\Lambda_A = \Lambda / 2 \lambda .
\end{equation}
With $t_j = \pi 2^j$ we have
\begin{align}
	N(j) & = 8 L^2  \left( \frac{2 \pi^3 \Lambda_A^3 8^j }{\epsilon_j}  \right)^{\frac{1}{2}}.
\end{align}	
The optimal choice of $\epsilon_j$ obeying the relevant constraints is again
\begin{equation}
	\epsilon_j = \epsilon_{\mathrm{tot}} \frac{ 2^j }{ 2( 2^m -1) } \sim \epsilon_{\mathrm{tot}}  2^{j-1-m}  ,
\end{equation}	
so that
\begin{align}
	N(j) & =  8 L^2  \left( \frac{2^{m+2}  \pi^3 \Lambda_A^3 4^j }{\epsilon_{\mathrm{tot}}}  \right)^{\frac{1}{2}}
\end{align}	
This leads to
\begin{align}
	N  = \sum_{j=1}^m N(j) & = 8 L^2  \left( \frac{2^{m+1}  \pi^3 \Lambda_A^3  }{\epsilon_{\mathrm{tot}}}  \right)^{\frac{1}{2}} \sum_{j=1}^{m} 2^j \\ \nonumber
	 & = 8 L^2  \left( \frac{2^{m+1}  \pi^3 \Lambda_A^3  }{\epsilon_{\mathrm{tot}}}  \right)^{\frac{1}{2}} 2(2^m - 1)  \\ \nonumber
	 	 & \sim 8 L^2  \left( \frac{2  \pi^3 \Lambda_A^3  }{\epsilon_{\mathrm{tot}}}  \right)^{\frac{1}{2}}  2^{\frac{3}{2}(m+1)} \\ \nonumber
\end{align}	
Using Eq.~\eqref{Eq2m} we find
\begin{equation}
	2^{3m/2} = ( 2^m )^{3/2} =  \frac{\lambda^{3/2}}{2^{3/2} \delta_E^{3/2}}\left( \frac{1+p_f}{ p_f}  \right)^{3/2} .
\end{equation}	
and so
\begin{equation}
	2^{\frac{3}{2}(m+1)} =  \frac{\lambda^{3/2}}{\delta_E^{3/2}}\left( \frac{1+p_f}{ p_f}  \right)^{3/2} .
\end{equation}	
Substituting this in, we get
\begin{align}
	N  & \sim 8 L^2  \left( \frac{ \lambda^3 \pi^3 \Lambda_A^3  }{  \delta_E^3 } \right)^{\frac{1}{2}}  \left( \frac{1+p_f}{p_f \epsilon_{\mathrm{tot}}^{1/3} } \right)^{3/2} .
\end{align}
We define the contents of the second round pair of brackets as
where in the last line we define
\begin{align}
	 Y & :=  \frac{1+p_f}{p_f \epsilon^{1/3} }  \\ \nonumber
	  & = 2^{1/3}  \frac{1+p_f}{p_f (P_f -p_f)^{1/3} }  .
\end{align}	
Again, we minimise this, assuming constant $P_f$.  We find that for small $P_f$, the optimal is given by choosing $p_f = (3/4) P_f$.  This leads, in the small $P_f$ limit, to $Y^{3/2} \sim 4.35 / P_f^{2}$ and therefore
\begin{align}
	N  & \sim (8 * 4.35) L^2  \left( \frac{ \lambda^3 \pi^3 \Lambda_A^3  }{  \delta_E^3 } \right)^{\frac{1}{2}}  \frac{1}{P_f^2} .
\end{align}
Using Eq.~\eqref{EqlambdaA} we get
\begin{align}
	N  & \sim \frac{ (8 * 4.35) \pi^{3/2}}{\sqrt{8}} L^2   \frac{   \Lambda^{3/2}  }{  \delta_E^{3/2}  P_f^2 } \\
		 & =(\sqrt{8} * 4.35) \pi^{3/2}L^2   \frac{   \Lambda^{3/2}  }{  \delta_E^{3/2}  P_f^2 }  .
\end{align}
Evaluating the constant and rounding to nearest integer, we get
\begin{align}
	\label{Trotterphase}
	N  & \sim 69    \frac{  L^2 \Lambda^{3/2}  }{  \delta_E^{3/2}  P_f^2 }  .
\end{align}

\subsection{Comparison}

In Fig.~\ref{Fig_PhaseCosts} we plot Eq.~\eqref{qDRIFTphase} for qDRIFT and Eq.~\eqref{Trotterphase} for 2nd order Trotter, as an upper bound for the gate counts to implement phase estimation.  Our earlier numerics have already shown that higher order Trotter is not competitive in the relevant parameter regime.   At $P_f=5\%$ we see speedups of $\times 1406$, $\times 304$ and $\times 789$, respectively. This advantage decreases with smaller $P_f$ and vanishes around $P_f \sim 10^{-4}-10^{-5}$.  However, phase estimation always needed repetition when applied to a state that is not exactly the groundstate (see Sec.~\ref{SecFailureProbs}).  Therefore, as we have already argued, a modest failure probability $P_f = 10 \%-5\%$ is reasonable.  

We finish by repeating our earlier caveats that these plots show known rigorous upper bounds and that actual performance is expected to be many orders of magnitude better. It is even plausible that 2nd order Trotter regains the advantage when we consider actual performance.  The question of actual performance is difficult and beyond our present scope, but a clear direction for future work.  Furthermore, for clarity we considered an early proposal for phase estimation but more modern techniques would also significantly improve performance for both protocols.

\end{document}